\documentclass{article}

\usepackage{epsfig}
\usepackage{amsmath}
\usepackage{amssymb}
\usepackage{theorem}
\theorembodyfont{\upshape}

\newtheorem{theorem}{Theorem}

\title{General non-existence theorem for phase transitions in 
one-dimensional systems with short range interactions, and physical
examples of such transitions}

\author{Jos\'e A.\ Cuesta\thanks{\tt cuesta@math.uc3m.es}
\ and Angel S\'anchez\thanks{\tt anxo@math.uc3m.es}\\
\ \\
Grupo Interdisciplinar de Sistemas Complejos (GISC)\\
and Departamento de Matem\'aticas\\
Universidad Carlos III de Madrid\\
Avenida de la Universidad 30\\
28911 Legan\'es, Madrid\\
Spain\\ 
\ \\ 
{\tt http://gisc.uc3m.es}
} 

\begin{document}

\maketitle \thispagestyle{empty}

\begin{abstract}

We examine critically the issue of phase transitions in one-dimensional
systems with short range interactions. We begin by reviewing in detail the
most famous non-existence result, namely van Hove's theorem, emphasizing
its hypothesis and subsequently its limited range of applicability. To 
further underscore this point, we present several examples of one-dimensional
short ranged models that exhibit true, thermodynamic phase transitions,
with increasing level of complexity and closeness to reality. Thus having
made clear the necessity for a result broader than van Hove's theorem, we
set out to prove such a general non-existence theorem, widening largely 
the class of models known to be free of phase transitions. The theorem 
is presented from a rigorous mathematical point of view although examples
of the framework corresponding to usual physical systems
are given along the way. We close the paper with a discussion in more 
physical terms of the implications of this non-existence theorem. 

\bigskip

{\bf Running head:} Phase transitions in short-ranged 1D systems 

\bigskip

\end{abstract}

\newpage 

\pagestyle{plain}

\section{Introduction}

One-dimensional (1D) systems are among the most important and fruitful 
areas of research in Physics. This is due to the fact that such models
are generally much more amenable to analytical calculations than 
higher-dimensional ones, while describing to a certain degree many 
problems of actual physical relevance. Indeed, exact results for
1D systems have offered deep insights about very many phenomena 
which subsequently have led to advances in much broader
contexts. Remarkably, in spite of the large body of knowledge already
available about this class of problems, 1D systems still are a
continuous source of exciting new physics \cite{lieb:1966,bernasconi:1981}.
This is so in spite of the unjustifiable
prejudices or careless generalizations that prevent researchers 
from considering many 1D problems on the grounds of their lack 
of interest. This has been the case, for instance, with Anderson
localization in 1D disordered systems: although all proofs
in the literature are model dependent, for almost thirty
years it has been regarded as a general dictum that any kind and
amount of disorder will localize all electronic states in 1D.
During this time, almost no
researchers have studied localization in 1D as the previous statement
amount to consider it a case closed. However, thanks to a few works 
carried out with critical attitude, we now know that in the presence of
short-ranged \cite{dunlap:1990} or long-ranged {\em correlated} disorder
\cite{moura:1998} bands of extended states do exist. Subsequently, the 
breakdown of the belief on the generality of 1D localization 
phenomena has paved the way to most relevant results, such as, e.g., the
dependence of the transport properties of DNA on their information 
content (\cite{carpena:2002}; see also a partial retraction that does not
affect the DNA part of the paper in \cite{carpena:2003}).

In this paper, we undertake the critique of another famous general 
statement, namely that there cannot be phase transitions in 1D systems
with short range interactions.
This assertion is practically never questioned (see \cite{evans:2000} for 
a recent exception), even though no general
proof of it has ever been provided, an impossible task in view that
counterexamples have
been given more than thirty years ago as we will see below.
The influence of this piece of received
wisdom cannot be underestimated, and for the past fifty years has become
an almost unsurmountable barrier for any research on 1D phase transitions.
It is important then to remind the physics community of the limits of 
applicability of this result. To this end, we need to make the statement
rigorous for the widest possible
class of models. In doing so, (quasi) 1D physical systems exhibiting 
phase transitions will be again available for a host 
of applications; in addition, the possibility of using 1D models, often
exactly solvable ones, to advantageously study phase transitions will
be reopened. 

To carry out this program, we proceed along two complementary directions. 
First, we review the existing results about non existence of phase 
transitions in short-ranged 1D systems. To our knowledge, these
amount to a theorem proven by van Hove \cite{vanhove}
for homogeneous fluid-like
models, with pairwise interactions with a hard core and a cutoff,
and in the absence of an external field, which was later
generalized by Ruelle \cite{ruelle} to lattice models. We note that
the well known argument by Landau \cite{landau} about domain 
walls is heuristic and relies on approximate calculations; 
we will also comment briefly on this below. 
We then discuss several 1D models proposed in the past which exhibit
true thermodynamic phase transitions; these models have different degrees of 
complexity and closeness to physical situations, and we will pay special 
attention to the specific reasons why each of them is not 
included in the existing
theorems. Having thus established clearly the existence of phase 
transitions in 1D systems with short range interactions, we move to 
our second contribution, introducing rigorously 
a very general theorem on the
impossibility of phase transitions in such models. As we will see, the
theorem, which includes van Hove's and Ruelle's results as particular
cases, gives sufficient but not necessary conditions to forbid phase
transitions. We will also show how models not fulfilling one of the 
hypotheses exist which do have phase transitions and comment on the
ways to violate those hypotheses. Finally, we 
conclude the paper with a discussion focused on the physics underlying
the mathematical results presented.

\section{van Hove's theorem} 
\label{sec2}

When one encounters the sentence ``1D systems with short range interactions
can not have phase transitions'' in the literature, it is either considered
public knowledge and not supported by a quotation, or else is directly or
indirectly referred to a 1950 paper by van Hove, written in French
\cite{vanhove}. Indeed, the above statement
often receives the name ``van Hove's theorem''. However, there is nothing 
that general in the excellent work by van Hove, nor does he intend to 
mean it in his writing. It is very illuminating to quote the English 
abstract here: 

\begin{quotation}
``The free energy of a one-dimensional system of particles is calculated 
for the case of non-vanishing incompressibility radius of the particles
and a finite range of the forces. It is shown quite generally that no 
phase transition phenomena can occur under these circumstances. The 
method used is the reduction of the problem to an eigenvalue problem.''
\end{quotation}

Let us expand some more on the abstract, in order to understand exactly
what van Hove proved. He considered a system of $N$ identical
particles, lying on a segment of length $L$ on positions $x_i$,
$i=1,\ldots,N$, $0\leq x_i\leq L$. The potential energy of the system is
given by 
\begin{equation}\label{vanhove1} 
V=\sum_{i=1,i<j}^N U(|x_i-x_j|),
\end{equation}
with 
\begin{equation}\label{vanhove2} 
U(\xi)=\left\{\begin{array}{ll} +\infty, & \mbox{ if $0\leq\xi\leq d_0$,} \\
0, & \mbox{\ if $\xi\geq d_1$,}\end{array}\right.
\end{equation}
and $0<d_0<d_1$. 
We are thus faced with a system of hard-core
segments of diameter $d_0$, that interact only at distances smaller
than $d_1$; van Hove's remaining assumption about the interaction is
that $U$ is a continuous, bounded below function. 

The way he proves this result is, as he himself says, by reducing the 
problem to an eigenvalue problem. He is able to write the partition 
function of the system in terms of a transfer operator, whose largest
eigenvalue gives the only relevant contribution to the free energy in
the thermodynamic limit. After transforming the operator into a more
useful form, van Hove resorts to the theory of 
Fredholm integral operators and other theorems of functional analysis
to show that this eigenvalue is an analytic function of temperature 
and, consequently, that the system can not have phase transitions, 
understood rigorously as nonanalyticities of the free energy. The 
mathematical basis of this result will be made clear by the theorem
we will present later in this article, and therefore we do not need
to go into further
detail at this point (other than enthusiastically 
referring the interested reader to 
the original paper \cite{vanhove}). For the time being, suffice it to
say that the basic idea is an extension of the well-known Perron-Frobenius
theorem for non-negative matrices \cite{meyer,horn}; we will come back
to this theorem when discussing our first example in the next section.

The key point we want to make here relates to the {\em hypotheses 
needed to prove van Hove's theorem, i.e., to the class of systems
to which it applies}. Let us consider them separately: \vspace*{-2mm}

\paragraph{Homogeneity.}
First of all, the system has to be perfectly 
homogeneous, made up of {\em identical} particles. This automatically
excludes any inhomogeneous model, where inhomogeneous means either 
aperiodic or disordered. Periodic systems could in principle be 
included in the frame of van Hove's theorem by analyzing the transfer 
operator for a unit cell. This is a very strong restriction, and 
it should be very clear that any degree of inhomogeneity in the 
system makes it impossible to exclude phase transitions on the 
ground of van Hove's result.

\paragraph{No external fields.} van Hove's choice for the 
potential energy does not include terms depending on the position of
the particles $x_i$ alone, i.e., they only depend on relative 
interparticle distances. The simplest way to have those terms in the
potential is by introducing external fields. With such an addition
the model does not satisfy the hypothesis of van Hove's theorem and
might therefore have phase transitions.

\paragraph{Hard-core particles.} We do not need to insist much on the
finite range of the interaction potential, as this is almost always 
included in any statement about the impossibility of phase transitions
in 1D systems. It is much less known, however, that the validity of 
van Hove's result requires a hard core potential as well, meaning that
it does not apply to point-like or soft particles. 

\bigskip

Of these three conditions, the theorem we will introduce below will 
relax very much the second and third restrictions, although we will 
also present counterexamples showing that the theorem cannot be
extended to include any external field. Our work leaves open 
the question as to 
the types of external fields that may give rise to a phase
transition. As for the first condition, however,
we will say nothing about the inhomogeneous 
case. This is a much more complicated question, far beyond the 
scope of the present work, and that is why we want to stress here that
there is no known theorem forbidding phase transitions in 1D 
inhomogeneous systems. As a matter of fact, their existence is largely
acknowledged within the community working on the so called ``2D wetting'' 
on disordered substrates \cite{2Dwet}, 
a phenomenon described by inhomogeneous 1D models. 

To conclude this section, a comment is in order about extensions and
generalizations of van Hove's theorem. The most relevant one is due
to Ruelle \cite{ruelle,ruelle2}, who proved the lattice version of the theorem
under the same basic hypotheses (earlier, Rushbrooke and Ursell proved
it for the lattice gas with finite neighbor interaction \cite{rushbrooke}).
As for the finite range of the interactions, the work of Ruelle \cite{ruelle2}
and Dyson \cite{dyson} proved that pair interactions decaying as $1/r^2$ 
($r$ being the distance between variables) represent the boundary between 
models with and without phase transitions. Subsequently,
Fr\"ohlich and Spencer \cite{frohlich}
 showed that case $1/r^2$ was to be included in
those with phase transitions. We do not know of further results in 
this direction, and therefore this is as much as can be safely said about 
systems having or not having phase transitions in 1D. 

\section{Examples of 1D models with phase transitions} 

After reviewing the available results about non-existence of phase 
transitions in 1D systems with short-range interactions, we now 
present some selected examples where there indeed are true thermodynamic
phase transitions in spite of their 1D character and the range of 
their interactions. We proceed in order of difficulty, and try to 
cover the three main levels of transfer operators: finite matrices, 
infinite matrices and integral operators. Our first example is actually
very simple, and will allow us to review the transfer matrix formalism.
Both this one and the second model are exactly solvable, and will make it 
clear that phase transitions are certainly possible. The third model 
can be written in terms of a transfer operator as well, but the corresponding
eigenvalue problem can only be solved numerically. 

\subsection{Kittel's model}

\subsubsection{Model definition}

The first system we consider was proposed by Kittel in 1969 \cite{kittel}, 
and is closely related to another one introduced by Nagle a year earlier
\cite{nagle} as a simple model of KH$_2$PO$_4$ (usually known as KDP),
which exhibits a first-order phase transition as well. Incidentally, both 
papers were published in {\em American Journal of Physics}, which points
to the very pedagogical character of these models. Kittel's model is in
fact a single-ended zipper model, discussed ``as a good way to introduce
a biophysics example into a course on statistical physics'', and 
inspired in double-ended zipper models of polypeptide or DNA molecules. 

Kittel's model is as follows. Let us consider a zipper of $N$ links
that can be opened {\em only from one end}. If links 1, 2, $\ldots$, $n$ 
are all open, the energy required to open link $n+1$ is $\epsilon$;
however, if not all the preceding links are open, the energy required to 
open link $n+1$ is infinite. Link $N$ cannot be opened, and the zipper
is said to be open when the first $N-1$ links are. Further, we suppose 
that there are $G$ orientations which each open link can assume, i.e.,
the open state of a link is $G$-fold degenerated. As we will see below,
there is no phase transition if $G=1$, whereas for larger degeneracy 
a phase transition arises. In \cite{kittel} the partition function is
expressed as a geometric sum which can be immediately obtained, and 
subsequently all the magnitudes of interest can be calculated as well. 
Nevertheless, in order to introduce the context of this work, namely 
the transfer operator formalism, we will solve Kittel's model in terms
of a transfer matrix (Kittel's way is much simpler, see \cite{kittel},
but it is not a general procedure).
To this end, let us write the model Hamiltonian as
\begin{equation}
\label{kittel-ham}
{\cal H}_N=\epsilon\big(1-\delta_{s_1,0}\big)+\sum_{i=2}^{N-1}
\big(\epsilon+V_0\delta_{s_{i-1},0}\big)\big(1-\delta_{s_i,0}\big)
\end{equation}
where $s_i=0$ means that link $i$ is closed, $s_i=1,2,\ldots,G$ means
that the link is open in one of the possible $G$ states, and 
$\delta_{s,s'}$ is the Kronecker symbol.
Note that Kittel's constraint on the zipper corresponds to the
choice $V_0=\infty$, and that we have also imposed the boundary
condition $s_N=0$ (the rightmost end of the zipper is
always closed). The partition function will then be given by
\begin{equation}
\label{kittel-z}
{\cal Z}_N=\sum_{\rm config.}\exp\left(-\beta
{\cal H}_N\right),
\end{equation}
with $\beta=1/k_BT$ being the inverse temperature and where the sum
is to be understood over all configurations of the variables $s_i$,
$i=1,\dots,N-1$.

\subsubsection{Transfer matrix solution}

The transfer matrix formalism to compute the partition function is a
well-known technique in equilibrium Statistical Mechanics that can be
found in most textbooks (see, e.g., \cite{stanley, huang, plischke}).
To implement this procedure, we rewrite the partition function as
\begin{equation}
\label{kittel-z2}
{\cal Z}_N=\sum_{\rm config.}e^{-\beta\epsilon\big(1-\delta_{s_1,0}\big)}
\prod_{i=1}^{N-2} e^{-\beta\epsilon(1-\delta_{s_{i+1},0})}\big[1+
\big(e^{-\beta V_0}-1\big)
\delta_{s_i,0}\big(1-\delta_{s_{i+1},0}\big)\big].
\end{equation}
From now on, we follow Kittel and let $V_0=\infty$, which implies
that $e^{-\beta V_0}=0$. We introduce the
transfer matrix ${\bf T}=(t_{s,s'})$, defined as
\begin{equation}
\label{kittel-tss}
t_{s,s'}=e^{-\beta\epsilon(1-\delta_{s',0})}\big[1-
\delta_{s,0}\big(1-\delta_{s',0}\big)\big],
\end{equation}
or in matrix form
\begin{equation}
\label{kittel-t}
{\bf T}=\begin{pmatrix}
1 & 0 & \cdots & 0 \\
1 & a & \cdots & a \\
\vdots & \vdots &  & \vdots \\
1 & a & \cdots & a
\end{pmatrix},
\end{equation}
where $a\equiv e^{-\beta\epsilon}$. It is very important to realize that
the constraint that link $s_{i+1}$ cannot be open (cannot take the values
1 or 2) if link $s_{i}$ is closed ($s_i=0$) yields the null entries in the
first row of ${\bf T}$.

The partition function can thus be recast in the form
\begin{equation}
\label{kittel-z3}
{\cal Z}_N=\begin{pmatrix} 1 & a & \cdots & a \end{pmatrix}
{\bf T}^{N-2} \begin{pmatrix} 1 \\ 1 \\ \vdots \\ 1 \end{pmatrix}.
\end{equation}
Matrix ${\bf T}$ has three different eigenvalues, 
namely $\lambda_1=Ga$, $\lambda_2=1$ and $\lambda_3=0$. The
eigenvectors of the two nonzero eigenvalues are, respectively,
\begin{equation}
\label{eigenvectors}
{\bf v}_1=\begin{pmatrix} 0 \\ 1 \\ \vdots \\ 1 \end{pmatrix}, \qquad
{\bf v}_2=\begin{pmatrix} 1-Ga \\ 1 \\ \vdots \\ 1 \end{pmatrix},
\end{equation}
so, if we express
\begin{equation}
\begin{pmatrix} 1 \\ a \\ \vdots \\ a \end{pmatrix} =
\frac{a(1-Ga)-1}{1-Ga}{\bf v}_1+\frac{1}{1-Ga}{\bf v}_2,    \quad
\begin{pmatrix} 1 \\ 1 \\ \vdots \\ 1 \end{pmatrix} =
\frac{-Ga}{1-Ga}{\bf v}_1+\frac{1}{1-Ga}{\bf v}_2,
\end{equation}
we arrive finally at
\begin{equation}
\label{kittel-fin}
{\cal Z}_N=\frac{1-(Ga)^N}{1-Ga}=
\frac{1-(Ge^{-\beta\epsilon})^N}{1-Ge^{-\beta\epsilon}}
\end{equation}
in agreement with Kittel's result \cite{kittel} or, alternatively,
\begin{equation}
\label{kittel-fin2}
{\cal Z}_N=\frac{1}{1-Ge^{-\beta\epsilon}}(-\lambda_1^N+\lambda_2^N)
\end{equation}
which is more suitable to our purposes, and shows the general structure 
of transfer matrix results: the partition function is expressed as a
linear combination of $N$th powers of the transfer
matrix eigenvalues. In the thermodynamic limit,
only the contribution of the largest eigenvalue remains, and we 
have, as $N\to\infty$, that the free energy is given by 
\begin{equation}
\label{kittel-fin3}
f\equiv\frac{1}{N}{\cal F}\equiv -\frac{1}{\beta N}\ln {\cal Z}_N=
-\frac{1}{\beta}\ln\max (\lambda_1,\lambda_2).
\end{equation}

\begin{figure}
\begin{center}
\noindent
\epsfig{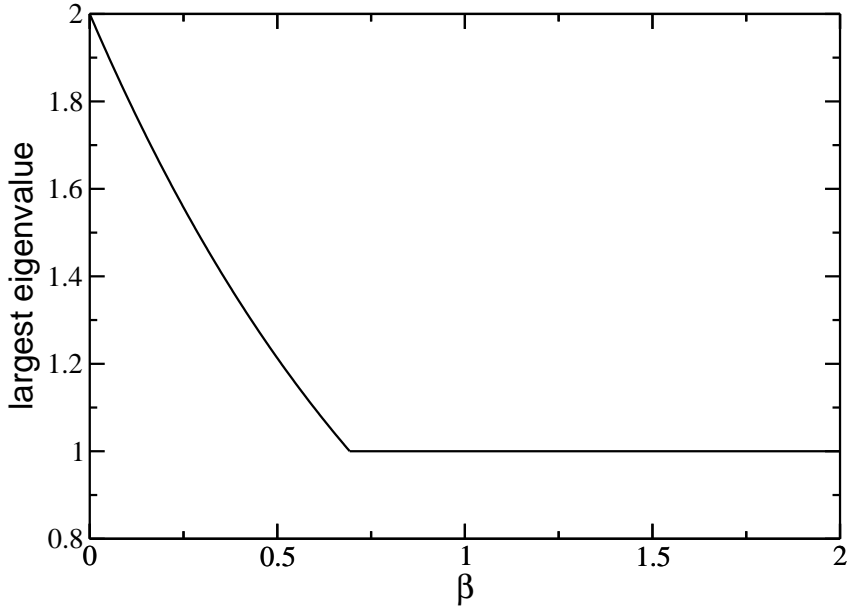}
\caption{Largest eigenvalue of the transfer matrix for Kittel's model
with $G=2$
vs inverse temperature, with $\epsilon=1$.
Note the nonanalyticity at $\beta=1/\ln 2$.}
\label{fig1}
\end{center}
\end{figure}

We are thus faced with the crux of the matter: in order to have a 
phase transition, meaning a nonanalyticity of the free energy---given
that the eigenvalues are positive, analytic functions of $\beta$---we
need two eigenvalues to cross at a certain
$\beta_c$. In our problem, we only have to compare $\lambda_1$
and $\lambda_2$ to find that they cross at a temperature given
by $\beta_c=\ln G/\epsilon$, or, equivalently,
$T_c=k_B\epsilon/\ln G$; above (below) $T_c$, $\lambda_1$
($\lambda_2$) is the largest eigenvalue (see Fig.\ \ref{fig1}).
At $T_c$, the derivative of the free energy is
discontinuous, implying that the specific heat diverges, i.e., we 
have a second order phase transition. 

It is interesting to note that $T_c=k_B\epsilon/\ln G$ is finite
as long as $G>1$; for the non-degenerate case $G=1$ (only one open
state) the transition takes place at $T=\infty$ or, in other words,
there is no phase transition.

\subsubsection{Discussion}

We are now in a position to explain in more detail the mathematical
reasons underlying these results as well as, generally speaking, 
van Hove's theorem on the absence of phase transitions. In the 
preceding section we mentioned that van Hove's theorem relies on an
extension of the Perron-Frobenius theorem for matrices to integral
operators; however, for our discussion of Kittel's model, we need
only the original result by Perron and Frobenius \cite{meyer,horn}:
\begin{theorem}[Perron-Frobenius]
Let ${\bf A}$ be a non-negative (all its elements are non-negative),
irreducible matrix; then its spectral radius (maximum eigenvalue)
$\rho({\bf A})>0$ is an eigenvalue of algebraic multiplicity one.
\label{th:perron-frobenius}
\end{theorem}
A matrix ${\bf A}$ is {\em irreducible} if there does not exist
a permutation matrix ${\bf P}$ such that 
\begin{equation}
{\bf P}^t {\bf A} {\bf P}=\left(\begin{array}{cc} {\bf X} & {\bf Y} \\
{\bf 0} & {\bf Z}\end{array}\right)
\end{equation}
with both ${\bf X}$ and ${\bf Z}$ being square submatrices. 

Let us note that this theorem is not enough for our purposes, because we 
are not dealing with a specific matrix; instead, we are considering
a family of matrices depending on temperature, ${\bf T}(\beta)$.
We also need the following result (see ref.~\cite{kato:1995}, sec.~II.1.8),
valid for matrices analytic in $\beta$ (all their elements are analytic
functions of $\beta$):

\begin{theorem}
For every $\beta$ in a simply connected set $D\subset\mathbb{C}$,
let ${\bf T}(\beta)$ be a linear operator on an $n$-dimensional
vector space $X$ (i.e.\ ${\bf T}(\beta)$ is an $n\times n$ complex
matrix). Assume that ${\bf T}(\beta)$ is analytic in $D$. Let $\Sigma$
be a subset of eigenvalues of ${\bf T}(\beta)$ whose number, $s$, remains
constant for all $\beta\in D$ (i.e., eigenvalue splitting does not 
occur). Then each eigenvalue of $\Sigma$ has
constant multiplicity and can be expressed as an analytic function in $D$,
$\lambda_j(\beta)$ ($j=1,\dots,s$).
\label{th:eigen-finite}
\end{theorem}
So for a non-negative, irreducible transfer matrix ${\bf T}(\beta)$ whose
elements are analytic functions in a neighborhood of the positive real 
axis, $\beta>0$, theorems \ref{th:perron-frobenius} and \ref{th:eigen-finite}
imply that the maximum eigenvalue (hence the free energy) is an analytic
function of $\beta$, for all $\beta>0$.

We can now turn to the reasons as to why there is a phase transition 
in Kittel's model. We stress that 
transfer matrices, made up from Boltzmann factors, i.e., exponentials,
are always strictly positive and, consequently, irreducible and analytic
in $\beta$. Under these conditions there cannot be a phase transition for
any finite $\beta>0$. Therefore, the only way we can escape the hypothesis
of the Perron-Frobenius theorem is by assigning an
infinite energy to some configurations, thus giving rise to null entries
in the matrix, which may or not then be irreducible. This is exactly 
the case in Kittel's model. It is important to realize that breaking 
the irreducibility hypothesis does not ensure eigenvalue crossing: 
Kittel's model transfer matrix for the non-degenerate case, $G=1$, is
also reducible, and the eigenvalue crossing takes place only at
$\beta=0$, as we have already explained, yielding the analyticity of the
maximum eigenvalue (hence of the free energy) for any finite temperature.

Summarizing, Kittel's model has allowed us to show how phase transitions
can take place in 1D models whose statistical mechanics can be computed
with $n\times n$ matrices or, equivalently, in 1D lattice models with 
a finite number of states per node and finite range of interactions. 
Due to the nature of the transfer matrix and the theorems that apply to
it, phase transitions are impossible (with the caveat about boundary 
conditions discussed in subsection \ref{subbc}) unless there are forbidden 
(infinite energy) configurations, but fulfilling this condition does not
necessarily induce a phase transition. As we will now see, this clear-cut
conclusion will become more and more complicated as transfer matrices 
of infinite size or integral
transfer operators are considered. The next two subsections 
will discuss briefly two such examples before proceeding to the 
detailed, rigorous discussion of the corresponding theorems. 

\subsection{Chui-Weeks's model}
\label{subcw}

We now proceed in order of increasing mathematical complexity and 
consider a model in which the transfer matrix has
infinite size. The specific example we consider was proposed by 
Chui and Weeks \cite{chui} and is given by the following Hamiltonian: 
\begin{equation}
\label{chui-h}
{\cal H}_N=J\sum_{i=1}^N|h_i-h_{i+1}|-K\sum_{i=1}^N\delta_{h_i,0}.
\end{equation}
This is a typical instance of the family of models called solid-on-solid
(SOS) for surface growth, in which $h_i$ stands for the height above
site $i$ of the lattice; the reason for the name SOS is that overhangs
are not allowed, i.e., the surface profile is single-valued. We will 
consider that heights can take on only integer values and that there
is an impenetrable substrate, imposing  $h_i\geq 0$. In this
context, the first term represents the contribution of surface tension
to the total energy, and the second one introduces an energy binding
the surface to the substrate. As we will comment below, this is 
crucial for the model to exhibit a phase transition. Interestingly, 
these systems have often been considered as two-dimensional ones 
because of the fact that they represent interfacial phenomena on a
plane, and therefore they have been considered not relevant for the
1D phase transition issue. We stress here that the fact that $h_i$ 
stands for a height does not change the 1D nature of the model, as 
it could equally well represent any other magnitude or internal degree
of freedom, not associated to a physical dimension.

Instead of following Chui and Weeks's presentation, which is very simple
but does not lead to explicit results, we resort to an alternative 
derivation proposed as exercise 5.7 in Yeomans's textbook \cite{yeomans}.
We will not go here into the details of the derivation and quote only
its main steps. A transfer matrix for the model is evidently
\begin{equation}
\label{chui-trans}
({\bf T})_{ij}\equiv e^{-\beta J|i-j|}\left[1+\left(e^{-\beta K}-1\right)
\delta_{i,0}\right], \quad i,j=1,2,\dots
\end{equation}
Note that the matrix dimension is actually infinite, as announced, and
stems from the fact that the amount of possible states (heights) at any
site of the lattice is infinite. It is
also important to realize that in this case none of the entries in the
matrix is zero, so we have a strictly positive matrix, although out of
the scope of the theorems discussed above because of its infinite dimension.

For simplicity, we introduce the notation $\omega\equiv e^{-\beta J}$,
$\kappa\equiv e^{-\beta K}$. Then, by considering eigenvectors of the
form 
\begin{equation}
\label{chui-eigenvec}
{\bf v}_q\equiv (\psi_0,\cos(q+\theta),\cos(2q+\theta),\ldots),
\end{equation}
it is a matter of algebra to show that there is a continuous spectrum
of eigenvalues, 
\begin{equation}
\label{chui-sigma}
\sigma({\bf T})=\left[\frac{1-\omega}{1+\omega},\frac{1+\omega}{1-\omega}\right].
\end{equation}
The phase transition arises because, in the range of temperatures such
that $\kappa>1/(1-\omega)$, there is an additional eigenvector, 
\begin{equation}
\label{chui-eigenvec2}
{\bf v}_0\equiv (\psi_0,e^{-\mu},e^{-2\mu},\ldots)
\end{equation}
with eigenvalue 
\begin{equation}
\label{chui-eigenval2}
\lambda_0=\frac{\kappa(1-\omega^2)(\kappa-1)}{\kappa(1-\omega^2)-1},
\end{equation}
which, when it exists, is the largest eigenvalue. Thus, we have found 
again another case of eigenvalue crossing in the transfer matrix, which
indicates the existence of a phase transition. The physics of the 
transition is that, for temperatures below $T_c$, the
temperature at which $\kappa=1/(1-\omega)$, the surface is bound to 
the substrate and henceforth is macroscopically flat; on the contrary,
above $T_c$ the surface becomes free and its width is unbounded. This 
is an example of the so called roughening (or wetting, depending
on the context) transitions. 

It is interesting to observe that, if the substrate is not impenetrable
and all integer values from $-\infty$ to $\infty$ are allowed for the
variables $h_i$, the transition disappears, and the surface is always
pinned to the line $h_i=0$ \cite{chui}, meaning that it is flat at 
all temperatures. As discussed by Chui and Weeks, this is closely 
related to the fact that, in Quantum Mechanics, a potential well always
have a bound state if it is located within the infinite line $[-\infty,
\infty]$, while it needs special parameters to have a bound state if
the well is at the left side of the semi-infinite line $[0,\infty]$. 
This comment will be in order later, when discussing the general 
theorem on the absence of phase transitions, because we will point 
out that the range of definition of the transfer operator can be 
crucial to suppress or to allow phase transitions. 

\subsection{Dauxois-Peyrard's model} 
\label{subdp}

We conclude this section on examples of phase transitions in 1D systems
with short range interactions by considering the situation in which 
the model is still defined on a lattice, but the variables at the lattice
sites are real valued. In this case, the infinite transfer matrix of the
previous subsection becomes an integral transfer operator, as we will 
see below. 
A good instance of this class of problems is the extension of the model
we have just discussed to real-valued heights, studied by Burkhardt 
\cite{burkhardt}. A transfer operator for Burkhardt's model is 
\begin{equation}
\label{burkhardt}
{\bf T}\phi(h)=\int_0^{\infty}\,dh'\,\exp\left[-\beta\left(J|h-h'|+
\frac{1}{2}(U(h)+U(h'))\right)\right]\,\phi(h'),
\end{equation}
where $U(h)$ is the potential well binding the surface to the substrate,
generalizing the Kronecker delta in Chui-Weeks's model. We will not
discuss Burkhardt's results in detail as they are qualitatively the 
same as in the discrete height version, including the suppression of 
the phase transition by considering a doubly infinite range for $h$. 
Let us simply point out that, in this case, the analogy with the 
quantum-mechanical problem, mentioned at the end of the previous 
subsection, of the existence of bound states in a 
1D well becomes exact, as the statistical mechanical problem can 
be mapped to a Schr\"odinger equation. 
We refer the interested reader to \cite{burkhardt} for details.

In order to include examples taken from different contexts, we want 
to discuss in this subsection a model for DNA denaturation, that, in
addition, is a much more realistic model than the toy model introduced
by Kittel and discussed in detail above. The model was proposed in 
\cite{dauxois} (see \cite{theodorakopoulos1} for recent results; 
see \cite{theodorakopoulos2} for a brief review on DNA denaturation 
models), and we will refer to it as Dauxois-Peyrard's model. The 
corresponding Hamiltonian is 
\begin{equation}
\label{dauxois-ham}
{\cal H}_N=\sum_{i=1}^N \left[\frac{1}{2}m\dot{y}_n^2+
D(e^{-\alpha y_n} -1)^2+W(y_n,y_{n-1})\right],
\end{equation}
where the variable $y_n$ represents the transverse stretching of the
hydrogen bonds connecting the two base pairs at site $n$ of the 
double helix of DNA (note that the molecule is supposed to be homogeneous).
The first term in the Hamiltonian is the kinetic energy, with $m$ being
the mass of the base pairs; the second term, a Morse potential, 
represents not only the hydrogen bonds between base pairs but also 
the repulsion between phosphate groups and solvent effects; finally,
the stacking energy between neighboring base pairs along each of the
two strands is described by the anharmonic potential
\begin{equation}
\label{dauxois-w}
W(y_n,y_{n-1})=\frac{K}{2}\left[1+\rho e^{-\alpha(y_n+y_{n-1})}\right]\,
(y_n-y_{n-1}^2).
\end{equation}
Once again, the partition function of the model can be written in terms
of an integral transfer operator, which in this case is given by 
[compare with Eq.\ (\ref{burkhardt})]
\begin{equation}
\label{dauxois-trans}
{\bf T}\phi(y)=\int_{-\infty}^{A}\,dx\,\exp\left[-\beta\left(W(y,x)+
\frac{1}{2}[V(y)+V(x)]\right)\right]\,\phi(x)
\end{equation}
where the upper limit in the integral, $A$, is a cutoff introduced for
technical reasons, but the limit $A\to\infty$ is well defined. 

The problem with the Dauxois-Peyrard model is that it is not possible
to solve exactly for the eigenvalues of the transfer operator. However, 
in \cite{dauxois} the combined use of analytical approximations and 
numerical computation of the eigenvalues allowed the authors to 
provide compelling evidence for a phase transition in the anharmonic
case [$\alpha\neq 0$ in the
Hamiltonian (\ref{dauxois-w})]. Indeed, their numerical 
results show, much as in the Chui-Weeks's model, a single eigenvalue
in the discrete spectrum that merges the band of the continuous 
spectrum at a finite temperature. The result agrees very well with 
numerical simulations of the model, showing that above the critical
temperature the double strand denaturates (i.e., the two strands
separate to a macroscopically large distance or, equivalently, the 
mean value of $y_n$ diverges), whereas below the critical temperature
the two strands remain bound. Most interestingly, the predictions of 
the model compare very well with experiments on short chains \cite{DNAexp}.
The authors claim \cite{theodorakopoulos1} that
van Hove's theorem does not apply here, among other reasons, because
of the presence of the external field term given by the Morse potential. 
Actually, we want to go beyond their claim and stress that van Hove's
theorem has nothing to do with this model, because it does not fulfill
other hypotheses as well (although, admittedly, the most noticeable 
violation is the external potential, which breaks the
required translation invariance).
Therefore, this phase transition should not 
be discussed in the framework of van Hove's theorem:
As we will see in the next section, the transfer operator is likely 
to be excluded of the more general theorem we will present, thus 
making it possible the existence of this phase transition. 

\section{A general theorem on the non-existence of phase transitions}

Once established the existence of phase transitions in one-dimensional
systems with finite-range interaction, we will consider the formulation
of an impossibility theorem sufficiently general as to include all
known particular cases of proven nonexistence of phase transitions
(namely, the theorems of van Hove \cite{vanhove}, Ruelle 
\cite{ruelle,ruelle2} and Perron-Frobenius \cite{meyer,horn}, at least).

Our guideline to look for such a generalization will be the Perron-Frobenius
theorem for nonnegative matrices. This theorem applies to homogenous lattice 
models in which the state variables defined on each node take on values
from a finite set (like e.g.\ Ising or Potts variables) and interact
only with a finite set of neighbors. The partition function of those
models can be defined in terms of the eigenvalues of a finite nonnegative
(its elements are Boltzmann's factors) transfer matrix, the kind of
object to which Perron-Frobenius theorem applies. But general 
one-dimensional models may differ from those lattice models in at least
one of two ways: they can be continuum models, and state variables
can take values on an infinite (either discrete or continuum) set.
In these cases the partition function can be expressed in terms of a
transfer operator on a certain infinite-dimensional linear space. Integral
operators or infinite matrices are two particular instances of such
operators, but they are not the only ones.

The problem to generalize Perron-Frobenius theorem to operators more general 
than finite matrices is to extend the notions of nonnegativeness and
irreducibility. This amounts to equip functional spaces with an order
which allows comparing functions (at least in certain cases). The theory
resulting from introducing order in Banach spaces and its consequences
for the spectral theory of linear operators defined on them has been
a topic of active research for mathematicians for quite some time
\cite{meyer-nieberg:1991,zaanen:1997}, and it is at the heart of this
realm where the desired extension is found.


\subsection{Mathematical background}

Much as the proof of non-existence of phase transitions in 1D lattice
models with finite-state variables interacting through a short-range
potential is based upon Perron-Frobenius theorem, that of general
1D models is based on a generalization of that theorem to a class
of transfer operators. Such a generalization, known as Jentzsch-Perron
theorem (a special case of which was employed by van Hove to obtain
his result \cite{vanhove}) reads as follows (the
present statement is a slightly simplified version of
Corollary 4.2.14 on p.~273 of \cite{meyer-nieberg:1991}):

\begin{theorem}[Jentzsch-Perron] Let $E$ be a Banach lattice and
${\bf T}\ne 0$ a linear, positive, irreducible operator in $E$. Assume
${\bf T}^k$ is compact for some $k\in\mathbb{N}$. Then its spectral
radius $\rho({\bf T})>0$
is an eigenvalue of ${\bf T}$ with multiplicity one.
\label{th:Jentzsch-Perron}
\end{theorem}

The proof of this theorem roots deeply into the theory of Banach
lattices. The interested reader is urged to study the specialized
literature \cite{meyer-nieberg:1991, zaanen:1997} to discover the rich
structure that order induces in ordinary Banach spaces. Instead
of that, we are simply giving here the necessary clues to make
this theorem a practical tool to investigate phase transitions
in models whose partition function can be written in terms of a
transfer operator.

\subsubsection{Banach lattices in a nutshell}

A vector space, $E$, is said to be an {\em ordered vector space}
if a partial order ($\leq$) is defined between its elements such that
if $f,g$ are elements of $E$, (i) $f\leq g$ implies $f+h\leq g+h$ for
any $h\in E$ and (ii) $f\geq 0$ implies $\alpha f\geq 0$ for every
$\alpha\geq 0$ in $\mathbb{R}$ (we then say that the order is
{\em compatible} with the vector space structure). If $(E,\leq)$
is also a {\em lattice} (a mathematical notion not to be confuse
with physical lattices), i.e.\ if for any $f,g\in E$, $\sup\{f,g\}$ and
$\inf\{f,g\}$ are in $E$, then we call $E$ a {\em Riesz space}.

In a real Riesz space it makes sense to define the {\em absolute value}
of a vector as $|f|=\sup\{f,-f\}$, because a Riesz space is a lattice.
The extension of this notion to complex Riesz spaces is $|f|=
\sup\{{\rm Re}\,(fe^{-i\theta}),\ 0\leq\theta<2\pi\}$ (notice that
the latter definition reduces to the former one for real elements of
the Riesz space). This element, though, is not guaranteed to belong
to the Riesz space or even to exist at all.

When a Riesz space $E$ has a norm, $\|\cdot\|$, such that for $f,g\in E$,
$|f|\leq |g|$ implies $\|f\|\leq \|g\|$ (i.e.\ {\em compatible}
with the order), then $E$ is a {\em normed Riesz space}.
If the normed Riesz space $E$ is complete in the
norm (i.e.\ every Cauchy sequence converges in $E$ or, in other
words, if $E$ is a Banach space), then $E$ is called a {\em Banach
lattice}. In a complex Banach lattice, completeness ensures that
$|f|$ (see above) is always a well-defined element of it.

\paragraph{Physically relevant examples of Banach lattices.}
The most common Banach spaces are also Banach lattices with the
natural order. For instance, $l^p$ ($1\leq p<\infty$), the
sequences $x=(x_n)_{n\in\mathbb{N}}$ of complex numbers with
$\sum_{n\in\mathbb{N}}|x_n|^p<\infty$, ordered componentwise
(i.e.\ for $x,y\in l^p$, $x=(x_n)_{n\in\mathbb{N}}$,
$y=(y_n)_{n\in\mathbb{N}}$, we say that $x\leq y$ if ${\rm Re}\,x_n
\leq{\rm Re}\,y_n$ and ${\rm Im}\,x_n\leq{\rm Im}\,y_n$).
We have the same property for spaces $L^p(X,\mu)$, the complex
functions on the point set $X$ (to be precise, the classes of
functions which are equal `almost everywhere') having
$\int_X|f|^p\,d\mu<\infty$. The order is then pointwise almost
everywhere, i.e.\ for $f,g\in L^p(X,\mu)$, $f\leq g$ if
${\rm Re}\,f(x)\leq {\rm Re}\,g(x)$ and ${\rm Im}\,f(x)\leq
{\rm Im}\,g(x)$ for all $x\in X$, except for a set of vanishing
$\mu$-measure.

\subsubsection{Linear operators on Banach lattices}
\label{sublin}

An important subset of a Riesz space is its {\em positive cone},
$E_+=\{f\in E:\ f\geq 0\}$. A linear operator, ${\bf T}:E\mapsto E$ is
said to be a {\em positive operator} if ${\bf T}E_+\subset E_+$. Every
positive operator in a Banach lattice is automatically (norm) bounded.
We say that one such linear operator is {\em irreducible} if the
only invariant ideals are $\{0\}$ and $E$. In short, a vector
subspace $A\subset E$ is an {\em ideal} of the Riesz space $E$
if for any $x\in A$ it contains all $y\in E$ such that $|y|\leq|x|$.
As we will show below, for some very common types of operators
there is a simpler characterization of irreducibility.

We say that ${\bf T}$ is a {\em compact operator} if it maps the unit ball
($\{x\in E:\ \|x\|\leq 1\}$) in a relatively compact set (one whose
closure is a compact set) of $E$. Compact operators are the closest
to finite matrices because of the very simple
structure of their spectra \cite{meyer-nieberg:1991}. The
continuous and residual spectra of compact operators are empty.
Also, every $\lambda\ne 0$ in the spectrum is an eigenvalue
of finite multiplicity. There is a finite or countable number of
eigenvalues, and if not finite, they can be arranged in a sequence
$(\lambda_n)_{n\in\mathbb{N}}$ such that $\lambda_n\to 0$ as
$n\to\infty$ ($\lambda=0$ may or may not be itself an eigenvalue).
Thus each $\lambda_n\ne 0$ is an isolated point in the spectrum. 
In general,
one of the easiest ways to prove that an operator is {\em not} compact is
showing that part of its spectrum is continuous.

\paragraph{Physically relevant examples of linear operators.}
In the Banach lattice $\mathbb{C}^n$ every linear operator (an
$n\times n$ complex matrix) is, of
course, compact. In $\ell^2$ a linear operator ${\bf T}$ can be
represented by an ``infinite by infinite'' matrix $(t_{ij})_{i,j
\in\mathbb{N}}$. A sufficient (not necessary)
condition for ${\bf T}$ to be compact is $\sum_{i,j\in\mathbb{N}}
|t_{ij}|^2<\infty$; or if ${\bf T}$ is of the special type that
$t_{ij}=0$ for $|i-j|>r$, for some fixed $r$ (a $2r+1$-diagonal
operator), then a necessary and sufficient condition for ${\bf T}$
to be compact is $\lim_{ij\to\infty}t_{ij}=0$ (i.e.\ every
diagonal is a null sequence) \cite{akhiezer:1993}. In $L^2(X,\mu)$,
an integral operator $({\bf T}f)(x)=\int_X t(x,y)f(y)\,d\mu_y$ with a
kernel $t(x,y)$ of the Hilbert-Schmidt type (i.e.\ with
$\int_{X^2}|t(x,y)|^2\,d\mu_xd\mu_y<\infty$) is compact.

For these particular classes of operators there are also simpler
tests of irreducibility.
In the case of $\mathbb{C}^n$ or $\ell^2$, ${\bf T}=(t_{ij})$ is
reducible if and only if there exists a finite nonempty subset
$A\subset\mathbb{N}$ such that
$\sum_{i\in A}\sum_{j\in A^c}t_{ij}=0$ ($A^c$ stands for the
complementary set of $A$). Likewise,
in the case of a Hilbert-Schmidt integral operator in
$L^2(X,\mu)$, ${\bf T}$ is reducible if and only if
there exists $A\subset X$ with $0<\mu(A)<\mu(X)$ such that
$\int_{A^c}\int_A|t(x,y)|^2\,d\mu_xd\mu_y=0$.
%

\subsubsection{Analyticity of the spectrum}

As in the case of finite matrices, there only remains to complete
this theorem with another one which guarantees the analyticity
of the maximum eigenvalue. Such a theorem is (from \cite{kato:1995},
sec.~VII.1.3):

\begin{theorem}
For every $\beta$ in a simply connected set $D\subset\mathbb{C}$,
let ${\bf T}(\beta)$ be a linear operator in a closed domain of a
Banach space $X$ (hence ${\bf T}(\beta)$ is a bounded operator).
Assume that ${\bf T}(\beta)$ is analytic in $D$ either in the strong or
in the weak convergence sense. Let $\Sigma$ be any finite set
of isolated eigenvalues of ${\bf T}(\beta)$
whose number of elements is constant in $D$.
Then each eigenvalue has constant multiplicity and can be
expressed as an analytic function in $D$, $\lambda_j(\beta)$
($j=1,\dots,|\Sigma|$).
\label{th:eigen-infinite}
\end{theorem}

\subsection{The theorem}

We are now in a position to formulate our result in precise terms.
Let us consider any statistical mechanical model whose partition 
function can be expressed as 
\begin{equation}
{\cal Z}_N=\varphi\left({\bf T}(\beta)^N\right),
\label{ZNfunctional}
\end{equation}
where ${\bf T}(\beta)$ is a transfer operator of any kind 
for every $\beta>0$, and $\varphi(\cdot)$ is a real, linear functional.
Typical instances of $\varphi$ are $\varphi({\bf T})=\mbox{tr}({\bf T})$,
or $\varphi({\bf T})=\langle f,{\bf T}g\rangle$ with $\langle\cdot,\cdot
\rangle$ a scalar product, as in Kittel's model [cf.\
eq.~(\ref{kittel-z3})], etc. Notice in passing that the partition
function of every 1D model with short-range interaction fit in
eq.~(\ref{ZNfunctional}), but not only. Models in $D>1$ can also
have a partition function given by eq.~(\ref{ZNfunctional}); the
only constraint is that ${\bf T}$ does not depend on $N$.

For such models we can now state the following theorem,
consequence of theorems \ref{th:Jentzsch-Perron} and
\ref{th:eigen-infinite}, which defines a class of models for
which there is {\em no} phase transition:

\begin{theorem}[Nonexistence of phase transitions]
Let ${\bf T}(\beta)$ be a compact, positive, irreducible, linear
operator on the Banach lattice $E$ for every $\beta$ in a complex
neighborhood containing $\beta>0$. Let $\lambda_{\rm max}(\beta)$ and
${\bf P}_{\rm max}(\beta)$ be, respectively, the maximum eigenvalue
of ${\bf T}(\beta)$ and the projector on its corresponding eigenspace.
Let $\varphi(\cdot)$ be a real, linear functional on the space of
bounded, linear operators on $E$ such that $\varphi\big({\bf P}_{\rm max}
(\beta)\big)\ne 0$. Then
\begin{equation}
\lim_{N\to\infty}\frac{1}{N}\ln{\cal Z}_N=-\ln\lambda_{\rm max}(\beta)
\label{eq:lnZlambdamax}
\end{equation}
is an analytic function on $\beta>0$, where ${\cal Z}_N$ is
given by eq.~(\ref{ZNfunctional}).
\end{theorem}
\emph{Proof:} Since ${\bf T}(\beta)$ is compact we know that its
spectrum is purely discrete and of the form
$\sigma\big({\bf T}(\beta)\big)=\big\{\lambda_{n}(\beta)\big\}_{n\in I}$,
where $I$ is a finite or countable set of indices. Zero may or not
be included, and, if $I$ is countable, the remaining eigenvalues can
be sorted in such a way that $\lambda_n\to 0$ as $n\to\infty$. Then
\begin{equation}
{\cal Z}_N=\sum_{n\in I}\lambda_n(\beta)^N\varphi\big({\bf P}_n(\beta)\big).
\end{equation}
Notice that if $I$ is countable, the above series will be convergent
for sufficiently large $N$. By factoring $\lambda_{\rm max}(\beta)$ out
of the series
\begin{equation}
{\cal Z}_N=\lambda_{\rm max}(\beta)^N\left[\varphi\big(
{\bf P}_{\rm max}(\beta)\big)+\epsilon_N\right], \quad
\epsilon_N=\sum_{n\in I'}\left(
\frac{\lambda_{n}(\beta)}{\lambda_{\rm max}(\beta)}\right)^N
\varphi\big({\bf P}_n(\beta)\big),
\end{equation}
where $I'$ is $I$ with the index corresponding to $\lambda_{\rm max}(\beta)$
removed. Equation (\ref{eq:lnZlambdamax}) simply follows from the fact that
\begin{equation}
\lim_{N\to\infty}\left[\varphi\big(
{\bf P}_{\rm max}(\beta)\big)+\epsilon_N\right]^{1/N}=1
\end{equation}
because $\varphi\big({\bf P}_{\rm max}(\beta)\big)\ne 0$.

Now, ${\bf T}(\beta)$ fulfills the hypothesis of theorem
\ref{th:Jentzsch-Perron}, thus $\lambda_{\rm max}(\beta)>0$ has
multiplicity one. Then taking $\Sigma=\big\{\lambda_{\rm max}(\beta)\big\}$
in theorem \ref{th:eigen-infinite} it follows that this eigenvalue
is an analytic function in $\beta>0$ and the proof is complete.
\hfill $\blacksquare$

\subsection{Discussion}

\subsubsection{Boundary conditions}
\label{subbc}

Among the hypotheses of the theorem, the only one whose significance may 
not be evident 
is $\varphi\big({\bf P}_{\rm max}(\beta)\big)\ne 0$. As stated in 
the proof, this is actually needed to show that the partition function 
can be written in terms of the maximum eigenvalue. Actually, the condition
is related to the choice of boundary conditions for the system. In the
examples mentioned above, 
$\varphi({\bf T})=\mbox{tr}({\bf T})$ arises from periodic boundary 
conditions, whereas 
$\varphi({\bf T})=\langle f,{\bf T}g\rangle$ arises from fixed boundary
conditions given by the two vectors $f$ and $g$. The condition is then
excluding boundary conditions that would suppress the eigenstates of 
the maximum eigenvalue as allowed states for the model. Otherwise
nothing can be said about the existence or not of phase transitions and,
in fact, they are possible: 
As an illustrative example, consider a transfer matrix 
for a three-state system of the form
\begin{equation}
\label{exbc}
{\bf T}\equiv 
\left(\begin{array}{rrr} 3 & 1 & 1 \\ 1 & b & 1 \\ 1 & 1 & b \end{array}
\right)
\end{equation}
This is a positive, irreducible matrix which, according to Perron-Frobenius
theorem, can not have a phase transition. However, the spectrum of this 
matrix is 
\begin{equation}
\label{exbc2}
\sigma({\bf T})=\left\{b-1, \frac{1}{2}\left(4+b\pm \sqrt{12-4b+b^2}\right)
 \right\}.
\end{equation}
Choosing now the boundary conditions to be given by an eigenvector
orthogonal to that of the 
maximum eigenvalue, $(4+b+\sqrt{12-4b+b^2})/2$, 
the hypothesis on the projector of the theorem will not be fulfilled.
One can easily check that
in that particular case, as there is a crossing of the second and 
third eigenvalues at $b=3$, the model has a thermodynamic phase 
transition even if it is described by a positive, irreducible 
matrix.  Of course, this occurs only for those specific boundary 
conditions, and in general the model will behave in the usual way. 
Admittedly, this is an 
academic example because if matrix (\ref{exbc}) were to
represent the transfer matrix of a physical system, 
both the energy of the first state and 
the boundary conditions (through the corresponding eigenvectors)
would be temperature dependent.
It is conceivable, though, that operators with such features
could arise in more realistic systems. In any event, it is clear 
that the hypotheses 
on the projector is needed to prevent pathological
situations like this one.

\subsubsection{Previous examples of phase 
transitions in the context of the theorem}
\label{subpre}

Once we have the general result on the absence of phase transitions above,
it is the time to address the issue as to the two examples of phase 
transitions discussed in subsections \ref{subcw} and \ref{subdp},
namely the Chui-Weeks's and the Dauxois-Peyrard's models. The fact that
they do not conform to the type of operator in the theorem is clear in 
view that both operators possess continuous spectrum, which as mentioned
in subsection \ref{sublin}, makes it impossible for them to be compact. 
However, in using this mathematical condition to show that some model 
is outside the range of applicability of the theorem one must consider 
several subtleties: 

\paragraph{Analytical calculations.} For the Chui-Weeks's model, the 
spectrum of the infinite transfer matrix is obtained analytically, and
therefore non-compactness is rigorously established. Nevertheless, this
needs not be the case in general. A good example is provided by the 
1D sine-Gordon model, thoroughly discussed in \cite{cuesta:2002} and
defined by the following Hamiltonian:
\begin{equation}\label{hamil}
\mathcal{H}=\sum_{i=1}^N\big\{\frac{J}{2}(h_{i-1}-h_i)^2+
V_0[1-\cos(h_i)]\big\}. 
\end{equation}
From the fact that the potential term is periodic in $h$, it follows by
Floquet-Bloch theorem that the spectrum of the corresponding transfer
integral operator is continuous \cite{tsuzuki:1988}, which would in turn
imply that the model does not fulfill the hypothesis of the theorem and
subsequently, it could exhibit phase transitions. Note that this does not
imply that it must exhibit a phase transition: indeed, in \cite{cuesta:2002}
it was shown that a suitable change of variables casts the operator in 
a form compatible with the theorem, thus establishing the impossibility
of phase transitions in this model (and in fact in a much wider class).
Interestingly, the same problem arises in van Hove's theorem \cite{vanhove};
van Hove writes first his general transfer operator in a non-compact 
form, but he is able to rewrite it as a compact operator and to prove 
his theorem. 
The difference with respect to the Chui-Weeks's model is that in this 
case it is possible to calculate the spectrum and prove that there is 
actually an eigenvalue crossing. These considerations indicate that 
non-compactness of the transfer operator merely excludes it from the 
theorem, but is not enough to say anything definite about the possibility
of phase transitions. 

\paragraph{Numerical calculations.} The situation is more complicated 
with the Dauxois-Peyrard model, where the spectrum cannot be computed 
analytically, and only numerical results are available. Resorting to 
numerical algorithms to study the spectrum of such an integral operator
implies several difficult issues. To begin with, there are two sources
of numerical error involved: the discretization of the integral and 
the truncation of the integration range. In some cases, such as the 
sine-Gordon model, the latter problem can be avoided by a change of 
variable, see \cite{saul1}; however, the former one cannot be cured. 
Further, when discretizing an integral operator such as the ones we
are discussing here, the result is {\em always} a finite matrix that
is necessarily positive and hence irreducible: i.e., it is subject to
the Perron-Frobenius theorem and cannot have singularities in the 
largest eigenvalue. Hence, all that one can see in a numerical calculation
of the spectrum of an integral operator is a possibly rapid, but anyway
smooth, change of the behavior of the largest eigenvalue. In fact, if
discontinuities are observed, they must come from the lack of precision 
of the computation, which leads to the vanishing of very small matrix 
elements that effectively yield the matrix reducible. It is very 
important then to complete the study of the eigenvalues with other 
quantities, preferably the eigenstates themselves. A good example 
of such an analysis is given in \cite{dauxois,theodorakopoulos1}, 
where the existence of a phase transition in the Dauxois-Peyrard's 
model is firmly established even if it cannot be rigorously proven.

\section{Conclusions}

In this paper, we have attempted to convey two main conclusions: First, 
there are true thermodynamic phase transitions in one dimensional systems
with short range interactions, in spite of the widespread belief on the
opposite; and second, we have provided a very general theorem 
about non-existence of those transitions. In this closing 
section we discuss both conclusions and their implications. 

To be sure, the existence of phase transitions in 1D systems
with short range interactions is not a new result. In this respect, 
what we have done here is to collect
and present within a unified framework a few, selected instances of 
such phase transitions, the earliest of which were proposed already in
the sixties.
In our opinion there are two main reasons which can explain why
part of the scientific community do not believe in its existence.
The first one is the fact that, indeed, {\em most} 1D systems
with short range interaction do not undergo a phase transition
(except maybe a zero or infinite temperature). Van Hove's 
rigorous result, Ruelle's extension to lattice models and the
most common exactly solvable examples of statistical physics
(Ising model, Potts model, etc.) seem to suggest this conclusion.
Landau's argument (not a theorem, as pointed out in the
introduction, and therefore applicable to a not well defined
class of models) reinforces this point of view. So far so good
because we are just describing the genesis of a reasonable
conjecture. The second reason, however, is not scientific. It
has recently been pointed out that a big deal of papers
contain cites which the authors have not read \cite{simkin}.
This is very obvious in the case of van Hove's work, which
you often see it cited as ``the proof'' of impossibility of 
phase transitions in 1D models with short range interactions,
referring to models having little or nothing in common with
the model van Hove deals with. This has spread the belief
that such a proof exists. We hope that the present work
helps to remedy this situation by tracing a neat boundary
between the 1D systems about which it can be actually proved
that there is no phase transition and those about which
nothing can be said.

A second point that we want to stress is that, even if we have discussed
just three basic examples, there are many more (and there will surely
be more to come). It is important to realize that whereas Kittel's model 
is largely academic, Chui-Weeks's and Dauxois-Peyrard's models are
relevant in physical situations of the importance of surface
growth/wetting 
and DNA denaturation, respectively. This means that they cannot be 
disregarded as ``academic, non realistic systems'' and that phase
transitions in 1D problems must be considered in their own right. 
Furthermore, the examples we have discussed represent 
three different stages in complexity of the model description in terms
of transfer operators: finite matrices, infinite matrices, and integral
operators. However, there are transfer operators that do not belong in 
any of these classes, such as the ones defined through the evolution 
of dynamical systems \cite{beck}. These are in principle much more 
difficult to tackle, but on the other hand they open new fields to the
study of 1D phase transitions. 

Moving now to the other result of the paper, the theorem presented here
is a very general result about non-existence
of phase transitions in 1D, short-ranged systems, and hence
it constitutes the chief original contribution of this work. Improving
on the starting point of van Hove's theorem, we have proven a rigorous
result valid for any system whose partition function can be written in
terms of a transfer operator independent on the system size. We want
to emphasize this formulation because it goes beyond the dimensionality
of the models, although we could as well define 1D models as those whose
transfer operator does not depend on the size. In any event, the theorem
presented here applies to a much wider class of problems than the original
van Hove's theorem,
as we have removed two of its main three limitations
discussed in section \ref{sec2}:
our result is valid for point-like particles and in the presence of
external fields.

Notwithstanding the considerations above on the virtues of the theorem
we have proven, it is most important to realize that it is not the
final answer to this issue yet.
One direction in which much work is needed
is to turn this result into an ``if-and-only-if''
theorem. Clearly, this is a very ambitious goal and, in addition, it 
might not even be reachable. In fact, the present result gives already 
some hints that this is the case. Indeed, compactness is needed to 
show that there cannot be phase transitions in 1D systems, but its 
absence does not imply anything, as there are models with non-compact
transfer operators with (Dauxois-Peyrard) and without (sine-Gordon)
phase transition. It can be argued at this point that the latter case
can finally be rewritten as a compact operator, but then the question
arises as to what is the class of ``apparent non-compact'' operators, 
i.e., non-compact operators that can be recast as compact. This is 
obviously not an easy question.
In this respect, it is interesting
to note that in the theory of dynamical systems a more
general class of transfer operators arises (quasi-compact operators),
whose spectral properties also allow to show the impossibility of
phase transitions (whatever this means for a dynamical system).
However, showing that an operator is quasi-compact without
resorting to determine its spectrum is far more difficult than
the already difficult task of proving compactness, and we know
of no instance of an equilibrium statistical-mechanical 1D system
described by one such operator. The reader interested in this
generalization can consult Refs.~\cite{ruelle:1978,baladi:2000}.
As for positiveness, we face the same
kind of problems: Kittel's model with non-degenerate open states is 
described by a non-positive, reducible 2$\times$2 matrix which does
not have a phase transition (rather, the transition temperature is 
infinite). It appears then that if an ``if-and-only-if'' version of the
theorem exists, it will need much refinement of the present hypotheses. 

Another comment that stems from the discussion in the previous paragraph
is that the theorem, being general and with clear-cut hypotheses, is not
very easy to apply. The case of systems with a finite number of states 
per site is well dealt with, and the consequence of Perron-Frobenius
theorem is that 
forbidden energy configurations are necessary in order to have a 1D 
phase transition in that case; otherwise, the corresponding finite 
matrix is always within the theorem applicability irrespective of any 
other ingredient of the model. However, as the complication of transfer
operators increases, it becomes more and more difficult to show 
whether or not they verify the hypotheses of the theorem. Among the 
three basic conditions, namely positiveness,  irreducibility, and
compactness, the case for the first two is again simpler, as in 
general irreducibility needs non-positivity and this is usually linked
to the existence of configurations with infinite energy. The problem 
arises with compactness, as, aside from the simplest operators, it is
not a trivial
task either to prove or to disprove it. As we have discussed 
in subsections \ref{sublin} and \ref{subpre}, the spectrum of the 
operator may be of help, but it does not provide a general tool. 
This is then the key point in characterizing operators to check for 
the possibility of phase transitions. 

Finally, it must be borne in mind that all the results and discussion 
in this paper relate to {\em homogeneous systems}. Of the three 
conditions for van Hove's theorem to apply mentioned above, this is
the only one we have not been able to remove, as the study of non-homogeneous
systems involves stupendous mathematical difficulties. At the level of
systems with a finite number of states per site, the theory of random
matrices might shed some light on the problem, although we have not 
been able to find guidance to this end among the available results. 
For more complex systems, with infinite matrices or integrals as 
transfer operators, this is a largely unknown territory. We referred 
in section \ref{sec2} to examples 
of true phase transitions in specific disordered systems \cite{2Dwet}
which grant that the problem is an interesting, physically relevant 
one, albeit one that needs much more effort. 

\section*{Acknowledgments} 

We want to thank Mar\'{\i}a Jos\'e Mu\~noz Bouzo for her invaluable
assessment in the mathematics of Banach lattices. We also want to
thank Sa\'ul Ares, Charles Doering, Michel Peyrard, 
Maxi San Miguel, Ra\'ul Toral and Chris van den Broeck
for helpful discussions
on the physical implications of these results. 
A preliminary report of this work was presented at the {\em FisEs '02}
meeting in Tarragona, Spain, and we benefited greatly from interactions
with quite a few of the participants. This work has been
supported by the Ministerio de Ciencia y Tecnolog\'{\i}a
of Spain through grants BFM2000-0004 (JAC) and BFM2000-0006 (AS).


\begin{thebibliography}{99}

\bibitem{lieb:1966} E.\ H.\ Lieb and D.\ C.\ Mattis, {\em Mathematical Physics
in One Dimension} (Academic Press, London, 1966). 
\bibitem{bernasconi:1981} J.\ Bernasconi and T.\ Schneider, {\em Physics in
One Dimension} (Springer, Berlin, 1981). 
\bibitem{dunlap:1990} D.\ H.\ Dunlap, H.\-L.\ Wu, and P.\ Phillips, Phys.\ 
Rev.\ Lett.\ {\bf 65}, 88 (1990). 
\bibitem{moura:1998} F.\ A.\ B.\ F.\ de Moura and M.\ L.\ Lyra, Phys.\ Rev.\ 
Lett.\ {\bf 81}, 3735 (1998).
\bibitem{carpena:2002} P.\ Carpena, P.\ Bernaola-Galv\'an, P.\ Ch.\ Ivanov,
and H.\ E.\ Stanley, Nature {\bf 418}, 955 (2002). 
\bibitem{carpena:2003} P.\ Carpena, P.\ Bernaola-Galv\'an, P.\ Ch.\ Ivanov,
and H.\ E.\ Stanley, Nature {\bf 421}, 764 (2003). 
\bibitem{evans:2000} M.\ R.\ Evans, Brazilian J.\ Phys.\ {\bf 30}, 42 (2000).
\bibitem{vanhove} L.\ van Hove, Physica {\bf 16}, 137 (1950) (reprinted
in \cite{lieb:1966}, p.\ 28).
\bibitem{ruelle} D.\ Ruelle, {\it Statistical Mechanics: Rigorous Results},
(Addison-Wesley, Reading, 1989).
\bibitem{landau} L.\ D.\ Landau and E.\ M.\ Lifshitz, {\em Statistical
Physics Part 1} (Pergamon, New York, 1980).
\bibitem{meyer} C.\ D.\ Meyer, {\em Matrix Analysis and Applied Linear
Algebra} (SIAM, Philadelphia, 2000).
\bibitem{horn} R.\ A.\ Horn and C.\ R.\ Johnson, {\em Matrix Analysis}
(Cambridge University Press, Cambridge, 1985).
\bibitem{2Dwet} G.\ Forgacs, J.\ M.\ Luck, Th.\ M.\ Nieuwenhuizen, and 
H.\ Orland, Phys.\ Rev.\ Lett.\ {\bf 57}, 2184 (1986); B.\ Derrida, V.\ Hakim,
and J.\ Vannimenus, J.\ Stat.\ Phys.\ {\bf 66}, 1189 (1992); G.\ Giguliarelli
and A.\ L.\ Stella, Phys.\ Rev.\ E {\bf 53}, 5035 (1996); P.\ S.\ Swain
and A.\ O.\ Parry, J.\ Phys.\ A {\bf 30}, 4597 (1997); T.\ W. Burkhardt, 
J.\ Phys.\ A {\bf 31}, L549 (1998).
\bibitem{ruelle2} D.\ Ruelle, Comm.\ Math.\ Phys.\ {\bf 9}, 267 (1968). 
\bibitem{rushbrooke} G.\ Rushbrooke and H.\ Ursell, Proc.\ Cambridge 
Phil.\ Soc.\ {\bf 44}, 263 (1948). 
\bibitem{dyson} F.\ J.\ Dyson, Comm.\ Math.\ Phys.\ {\bf 12}, 91 (1969). 
\bibitem{frohlich} J.\ Fr\"ohlich and T.\ Spencer, Comm.\ Math.\ Phys.\ {\bf
81}, 87 (1982). 
\bibitem{kittel} C.\ Kittel, Am.\ J.\ Phys.\ {\bf 37}, 917 (1969).
\bibitem{nagle} J.\ F.\ Nagle, Am.\ J.\ Phys.\ {\bf 36}, 1114 (1968).
\bibitem{stanley} H.\ E.\ Stanley, {\em Introduction to Phase Transitions
and Critical Phenomena} (Oxford University Press, Oxford, 1971). 
\bibitem{huang} K.\ Huang, {\em Statistical Mechanics} (Wiley, Singapore,
1987).
\bibitem{plischke} M.\ Plischke and B.\ Bergersen, {\em Equilibrium Statistical
Physics} (World Scientific, Singapore, 1994). 
\bibitem{chui} S.\ T.\ Chui and J.\ D.\ Weeks, Phys.\ Rev.\ B {\bf 23}, 
2438 (1981).
\bibitem{yeomans} J.\ M.\ Yeomans, {\em Statistical Mechanics of Phase
Transitions} (Oxford University Press, Oxford, 1992). 
\bibitem{burkhardt} T.\ W.\ Burkhardt, J.\ Phys.\ A {\bf 14}, L63 (1981).
\bibitem{dauxois} T.\ Dauxois, M.\ Peyrard and A.\ R.\ Bishop, 
Phys.\ Rev.\ E {\bf 47}, R44 (1993); T.\ Dauxois and M.\ Peyrard, 
Phys.\ Rev.\ E {\bf 51}, 4027 (1995).
\bibitem{theodorakopoulos1} N.\ Theodorakopoulos, T.\ Dauxois, and 
M.\ Peyrard, Phys.\ Rev.\ Lett.\ {\bf 85}, 6 (2000); T.\ Dauxois, 
N.\ Theodorakopoulos, and M.\ Peyrard, J.\ Stat.\ Phys.\ {\bf 107}, 
869 (2002).
\bibitem{theodorakopoulos2}  N.\ Theodorakopoulos, 
{\tt http://arxiv.org/abs/cond-mat/0210188} (2002). 
\bibitem{DNAexp} A.\ Campa and
A.\ Giansanti, Phys.\ Rev.\ E {\bf 68}, 3585 (1998).
\bibitem{meyer-nieberg:1991} P.\ Meyer-Nieberg, {\em Banach Lattices} 
(Springer, Berlin, 1991).
\bibitem{zaanen:1997} A.\ C.\ Zaanen, {\em Introduction to Operator 
Theory in Riesz Spaces} (Springer, Berlin, 1997).
\bibitem{akhiezer:1993} N.\ I.\ Akhiezer and I.\ M.\ Glazman, {\em 
Theory of Linear Operators in Hilbert Space} (Dover, New York, 1993).
\bibitem{kato:1995} T.\ Kato, {\em Perturbation Theory for Linear 
Operators} (Springer, Berlin, 1995).
\bibitem{cuesta:2002} J.\ A.\ Cuesta and A.\ S\'anchez, J.\ Phys.\ A 
{\bf 35}, 2373 (2002).
\bibitem{tsuzuki:1988} T.\ Tsuzuki and K.\ Sasaki, Prog.\ Theo.\ 
Phys.\ Supp.\ {\bf 94}, 73 (1988). 
\bibitem{saul1} S.\ Ares, J.\ A.\ Cuesta, A.\ S\'anchez, and R.\ Toral,
Phys.\ Rev.\ E {\bf 67}, 046108 (2003).
\bibitem{simkin} M.\ V.\ Simkin and V.\ P.\ Roychowdhury,
{\tt http://arxiv.org/abs/cond-mat/0212043} (2002).
\bibitem{beck} C.\ Beck and F.\ Schl\"ogl, {\em Thermodynamics of 
Chaotic Systems: An Introduction} (Cambridge University, Cambridge, 
1993), chapter 21. 
\bibitem{ruelle:1978} D.\ Ruelle, {\it Thermodynamic Formalism}
(Addison-Wesley, Reading, 1978).
\bibitem{baladi:2000} V.\ Baladi, {\it Positive transfer operators and
decay of correlations} (World Scientific, Singapore, 2000).


\end{thebibliography}
\end{document}